\documentclass[12pt]{iopart}
\usepackage{graphicx}
\def\nue{\nu_{e}}
\def\num{\nu_{\mu}}
\def\nut{\nu_{\tau}}
\def\nmnt{$\nu_{\mu}\leftrightarrow\nu_{\tau}$~}

\def\lsim{\lower.7ex\hbox{${\buildrel < \over \sim}$}}
\def\gsim{\lower.7ex\hbox{${\buildrel > \over \sim}$}}

\begin{document}

\title{Present Status of the K2K Experiment}

%\title{Present status of the K2K Experiment\footnote{Talk at 3rd Conference on Physics beyond the Standard Model (BEYOND2002), Oulu, Finland, June~2-7,~2002.}}

\author{Yuichi Oyama\footnote{Talk at 3rd Conference on Physics beyond the Standard Model (BEYOND2002), Oulu, Finland, June~2-7,~2002.\\
E-mail address:~{\tt yuichi.oyama@kek.jp};
~~URL:~{\tt http://neutrino.kek.jp/}\~{\tt oyama}}\\
for K2K Collaboration\footnote{ The K2K collaboration includes 149 physicists from KEK,
ICRR, Kobe, Niigata, Okayama, Tohoku, Tokai,
SUT, Kyoto, Boston, U.C.Irvine, Hawaii, LANL, SUNY, Washington, Warsaw, Chonnam,
Dongshin, Korea and SNU. The official Web page of K2K experiment
is {\tt http://neutrino.kek.jp/}}}

\address{Institute of Particles and Nuclear Studies, \\
High Energy Accelerator Research Organization (KEK), \\
Tsukuba, Ibaraki 305-0801 Japan}
\begin{abstract}
New results from the K2K (KEK to Kamioka) long-baseline neutrino-oscillation experiment
is presented
\end{abstract}

\section{Introduction}

The K2K experiment\cite{Nishikawa,detect,cpvio,moriond,cairo}
is the first long-baseline neutrino-oscillation experiment
with a distance of hundreds of km and
using an accelerator-based neutrino beam.
The nominal sensitive region in the neutrino-oscillation parameters
is $\Delta m^{2} > 3\times 10^{-3}$eV$^{2}$.
This covers the parameter region suggested by the atmospheric neutrino
anomaly observed by several underground experiments,\cite{Kamioka,IMB,Soudan}
and confirmed by Super-Kamiokande~(SK)\cite{SuperK}.

An overview of the K2K experiment is as follows.
An almost pure wide-band $\nu_{\mu}$ beam generated in the KEK
12-GeV Proton Synchrotron (PS) is detected in the SK
detector at a distance of 250 km.
Two beam monitors are located along the beam line.
One is a pion monitor (PIMON), which is a gas Cherenkov detector installed
downstream of the second magnetic Horn.
The other is a muon monitor (MUMON), which is an ionization chamber and
a silicon pad array behind the beam dump.    
In addition, two different types of
front detectors (FDs) are located at the KEK site.
One is a 1kt water Cherenkov detector~(1KT), which is
a miniature of the SK detector.
The other is a so-called fine-grained detector (FGD),
which is composed of a scintillating fiber tracker~(SFT)\cite{scifi},
trigger counters~(TRG), lead glass counters~(LG)
and a muon range detector~(MRD)\cite{mrd}.

Since the neutrino beamline, the beam monitors and detectors are already described
in previous articles\cite{cpvio,moriond,cairo}, they are not described
here. Instead, summaries of these detector components are given in tables.
Table~1 gives the design features of the neutrino beamline.
Table~2 gives the design and performance features of beam monitors.
The design, purpose, and  performance of the detectors are given in Table~3.
A schematic view of the Front detectors is also shown in Fig.1.

\begin{table}[t!]
\begin{center}
\caption{Design of the neutrino beam line.}
\begin{tabular}{ll}
\hline
\hline
~~$\bullet$ Proton kinetic energy  & 12 GeV\\
~~$\bullet$ Proton intensity       & 5.6 $\times 10^{12} \rm{proton/pulse}$\\
~~$\bullet$ Extraction mode        & Fast extraction\\
~~$\bullet$ Beam duration          & $\sim 1.1~\mu$second for every 2.2 second\\
~~$\bullet$ Target                 & Aluminum (3cm$\phi \times$ 65cm)\\
~~$\bullet$ Decay tunnel           & 200 m\\
\hline
\hline
\end{tabular}
\end{center}
\end{table}

\begin{table}[]
\begin{center}
\caption{List of the beam monitors along the K2K neutrino beamline.}
\begin{tabular}{ll}
\hline
\hline
\multicolumn{2}{l}{{\bf Pion monitor (PIMON)}}\\
~~$\bullet$ Design     & Gas Cherenkov detector with a spherical mirror and R-C318 gas. \\
                       & Reflected light is collected by 20 Hamamatsu R5600-01Q PMTs.\\
~~$\bullet$ Position   & Downstream of the second magnetic horn.\\
~~$\bullet$ Purpose    & Cherenkov photons from charged pions are measured\\
                       & at several refraction indices controlled by the gas pressure.\\
                       & The momentum and angular distributions of the pion beam are obtained\\
                       & from the intensity and shape of the Cherenkov image in the focal plane.\\
                       & The neutrino energy spectrum for $E_{\nu} > 1$GeV can be calculated\\
                       & from the kinetics of the pion decay.\\
\hline
\multicolumn{2}{l}{{\bf Muon monitor (MUMON)}}\\
~~$\bullet$ Design     & 2m $\times$ 2m ionization chamber filled with He gas\\
                       & and a silicon pad array.\\
~~$\bullet$ Position   & Downstream of the beam dump.\\
~~$\bullet$ Purpose    & The position of the beam center is monitored with an accuracy of\\
                       & less than 2 cm in spill on spill basis.\\
\hline
\hline
\end{tabular}
\end{center}
\end{table}

\begin{figure}[b!]
\center{
\caption{
Front detectors in the K2K experiment.
}
\includegraphics[height=8.0cm]{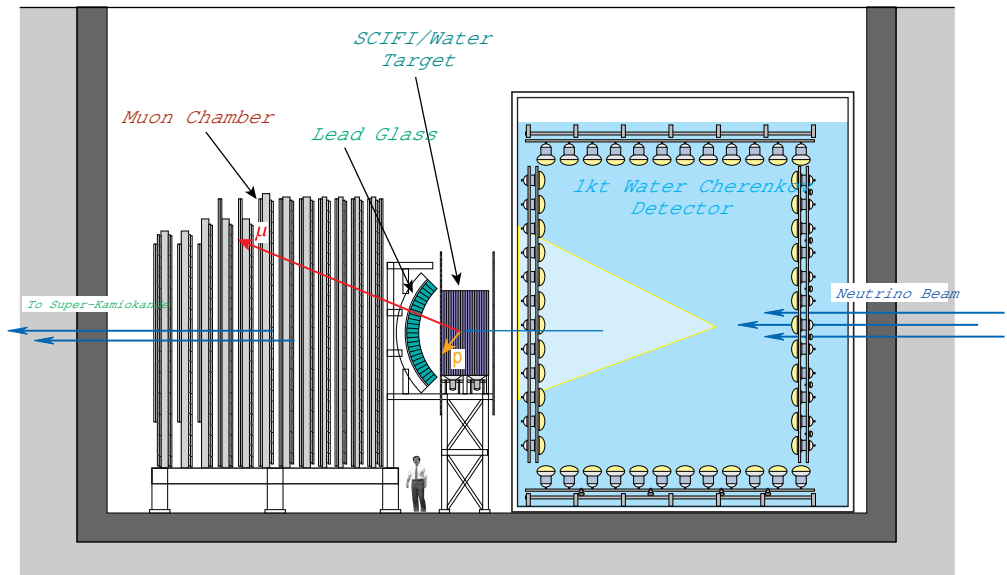}
\label{k2kfd}
}
\end{figure}

\begin{table}[p]
\begin{center}
\caption{Design, purpose, performance and event rate of the detectors used the K2K experiment.}
\begin{tabular}{ll}
\hline
\hline
\multicolumn{2}{l}{{\bf Front detectors 300m downstream}} \\
\multicolumn{2}{l}{{\bf (A)1kt water Cherenkov Detector}} \\
~~$\bullet$ Design     & A 1/50 (total volume) miniature of the Super-Kamiokande detector; 496 tons\\
                       & of water are viewed by 680 20-inch $\phi$ PMTs. The fiducial mass is 25.1~ton.\\
~~$\bullet$ Purpose    & Direct comparison with neutrino events in Super-Kamiokande\\
~~$\bullet$ Performance& $e/\mu$ identification capability:~~$>$99\%;~~~$\Delta E_{e} / E_{e}$:~~3\%/$\sqrt{E_{e}{\rm (GeV)}}$\\
~~$\bullet$ Event rate & $\sim$0.005 events/spill in fiducial volume; $\sim$1 $\times 10^{5}$ events for $10^{20}
$ p.o.t.\\

\multicolumn{2}{l}{{\bf (B)Fine Grained Detector}}\\
~~$\bullet$ Design     & Consists of a scintillating fiber tracker, trigger counters,\\
                       & lead-glass counters and a muon range detector.\\
~~$\bullet$ Purpose    & Precise measurement of the neutrino profile and energy distribution\\

\multicolumn{2}{l}{~{\bf (B-1)Scintillating fiber tracker (SFT)}}\\
~~~$\bullet$ Design   & 20-layer \char'134 sandwich'' of scintillating fiber (0.7mm$\phi$) sheets\\
                      & and water in aluminum containers.\\
                      & Sensitive area:~~ 2.4m $\times$ 2.4m; Fiducial mass:~~5.94 tons\\
~~~$\bullet$ Function & Track reconstruction of charged particles and identification of the kinematics\\
                      & of neutrino interactions. Water is used as the target material.\\ 
~~~$\bullet$ Performance & Detection efficiency of each layer:~~$>$99\%\\
                         & Position resolution:~~$\sim 0.6$mm\\
~~~$\bullet$ Event rate  & 0.001 events/spill, 2 $\times 10^{4}$ events for $10^{20}$ p.o.t.\\

\multicolumn{2}{l}{~{\bf (B-2)Trigger counters (TRG)}}\\
~~~$\bullet$ Design   & 80 large plastic scintillators (466cm (L)$\times$10.7cm(W)$\times$4.2cm(T)) \\
                      & covering upstream and downstream of the scintillating fiber tracker\\
~~~$\bullet$ Function & Rejection of cosmic ray muons and neutrino events from 1kt water\\
                      & Cherenkov detector, and measurement of the absolute event time.\\
~~~$\bullet$ Performance &  Timing resolution:~~$\sim$2nsec;~~~Position resolution:~~$\sim$5cm.\\
                         &  Detection threshold@center of the scintillators:~~$\sim$1.7MeV.\\
                         &  Detection efficiency for a penetrating charged particle:~~$>$ 99\%.\\

\multicolumn{2}{l}{~{\bf (B-3)Lead glass counters (LG)}}\\
~~~$\bullet$ Design   & 600 lead glass counters with an acceptance of 11.3cm $\times$12.2 cm each \\
~~~$\bullet$ Function  & Identification of electrons from the energy deposit in the counter\\
~~~$\bullet$ Performance &  $\Delta E_{e}/E_{e}$:~~$\sim 10\%/\sqrt{E_{e}{\rm (GeV)}}$.\\

\multicolumn{2}{l}{~{\bf (B-4)Muon range detector (MRD)}}\\
~~~$\bullet$ Design   & 12-layer \char'134 sandwich'' of $\sim$900 drift chambers\\
                      & and iron filters (10cm-20cm thickness).\\
~~~$\bullet$ Function  & Measurement of the muon energy from the range.\\
~~~$\bullet$ Performance &  Position resolution:~~2.2mm.\\
                         &  Detection efficiency of each layer:~~$\sim$ 99\%;~~~$\Delta E_{\mu}/E_{\mu}:~~8\sim 10\%$.\\
\hline
\multicolumn{2}{l}{{\bf Far detector (Super-Kamiokande) 250km downstream}}\\
~~$\bullet$ Design     & 50kt huge water Cherenkov detector at about 1000m underground.\\
                       & 22.5kt of the fiducial volume is viewed by 11164 20-inch $\phi$ PMTs.\\
~~$\bullet$ Performance& $e/\mu$ identification capability:~~$>$ 99 \%;\\
                       & $\Delta E_{e} / E_{e}$:~~3\%/$\sqrt{E_{e}{\rm (GeV)}}$;~~~$\Delta E_{\mu} / E_{\mu}$:~~3\%\\
                       & Accuracy of the absolute event time adjustment:~~$<$0.2$\mu$sec\\
~~$\bullet$ Event rate & $\sim$0.3 events/day, $\sim$160 events for $10^{20}$ p.o.t.\\
\hline
\hline
\end{tabular}
\end{center}
\end{table}

The K2K experiment was successfully started in early 1999, and physics data were
recorded from June 1999 to July 2001.
The total data-taking period during three years was 234.8~days.
The accumulated beam intensity was
$47.9 \times 10^{18}$ protons on the target (p.o.t.),
which is about 50\% of the goal of the experiment, $10^{20}$~p.o.t.

\section{Study of the Neutrino beam properties at the KEK site}
The characteristics of the neutrino beam at the KEK site
were examined using FDs and beam monitors.
In this section, the present status of analyses on (1)the neutrino beam direction,
(2)the neutrino beam intensity and its stability, (3)the $\nue/\num$ ratio,
(4)the neutrino energy spectrum, and
(5)extrapolation of the neutrino flux are presented.

\subsection{Neutrino beam direction}
The neutrino beam-line was constructed using a GPS position survey\cite{GPS};
the alignment of the beam-line, FDs and SK is better than 0.1~mrad.
The neutrino beam direction relative to the beam-line was measured
with MUMON and MRD independently. 

The position resolution
of MUMON is about 2 cm, corresponding to an angular resolution of 0.1 mrad.
Because $\num$ and muons originate in the same pion decay in the decay volume,
the $\num$ beam direction can be examined from the profile
center of the muon beam.
The time variation of the profile center is plotted in Fig.~2.
The direction of the muon beam agrees with the beam-line within 1~mrad.

\begin{figure}[b!]
\center{{\includegraphics[height=6.0cm]{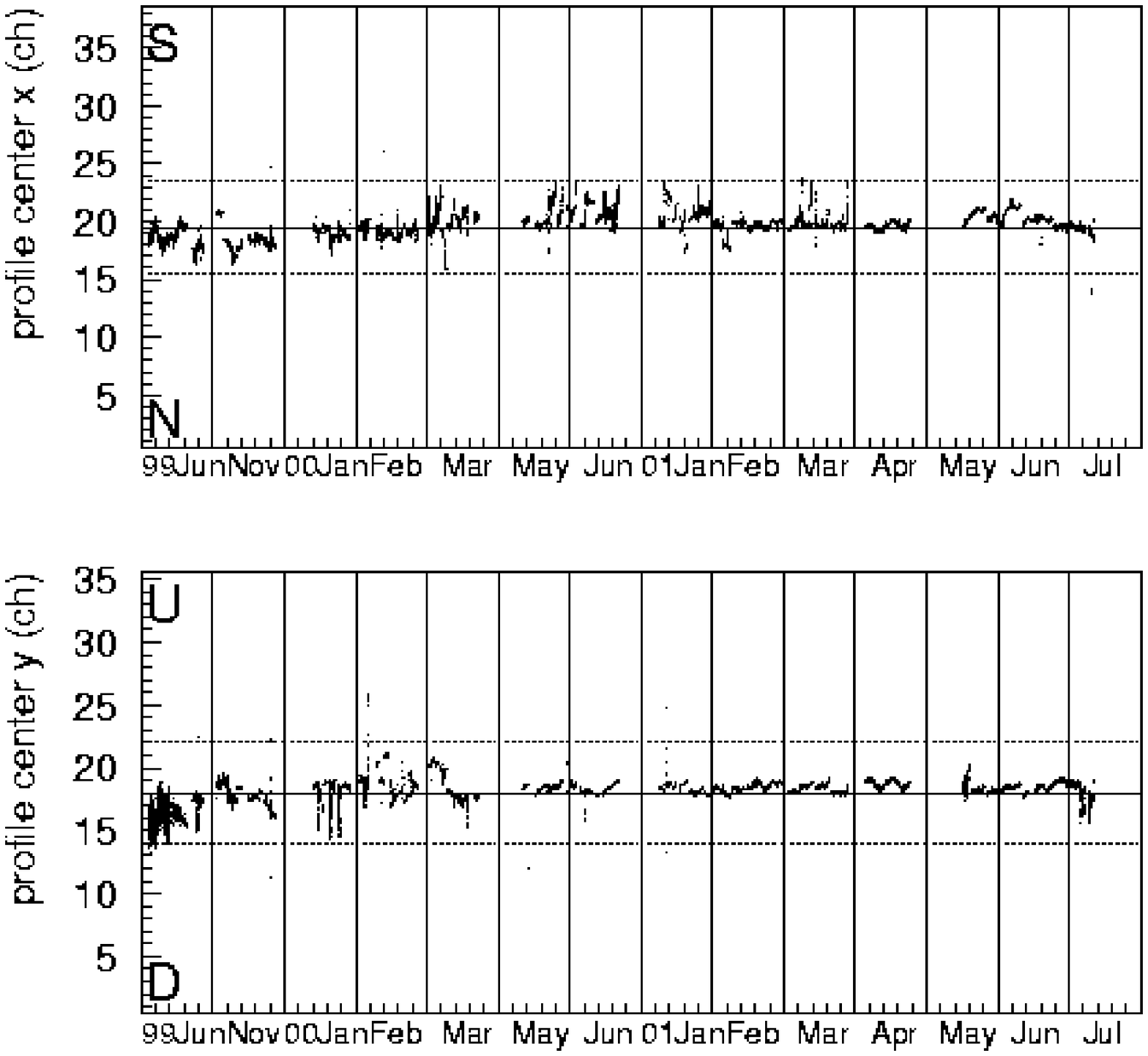}}
\hskip 5mm
        {\includegraphics[height=6.0cm]{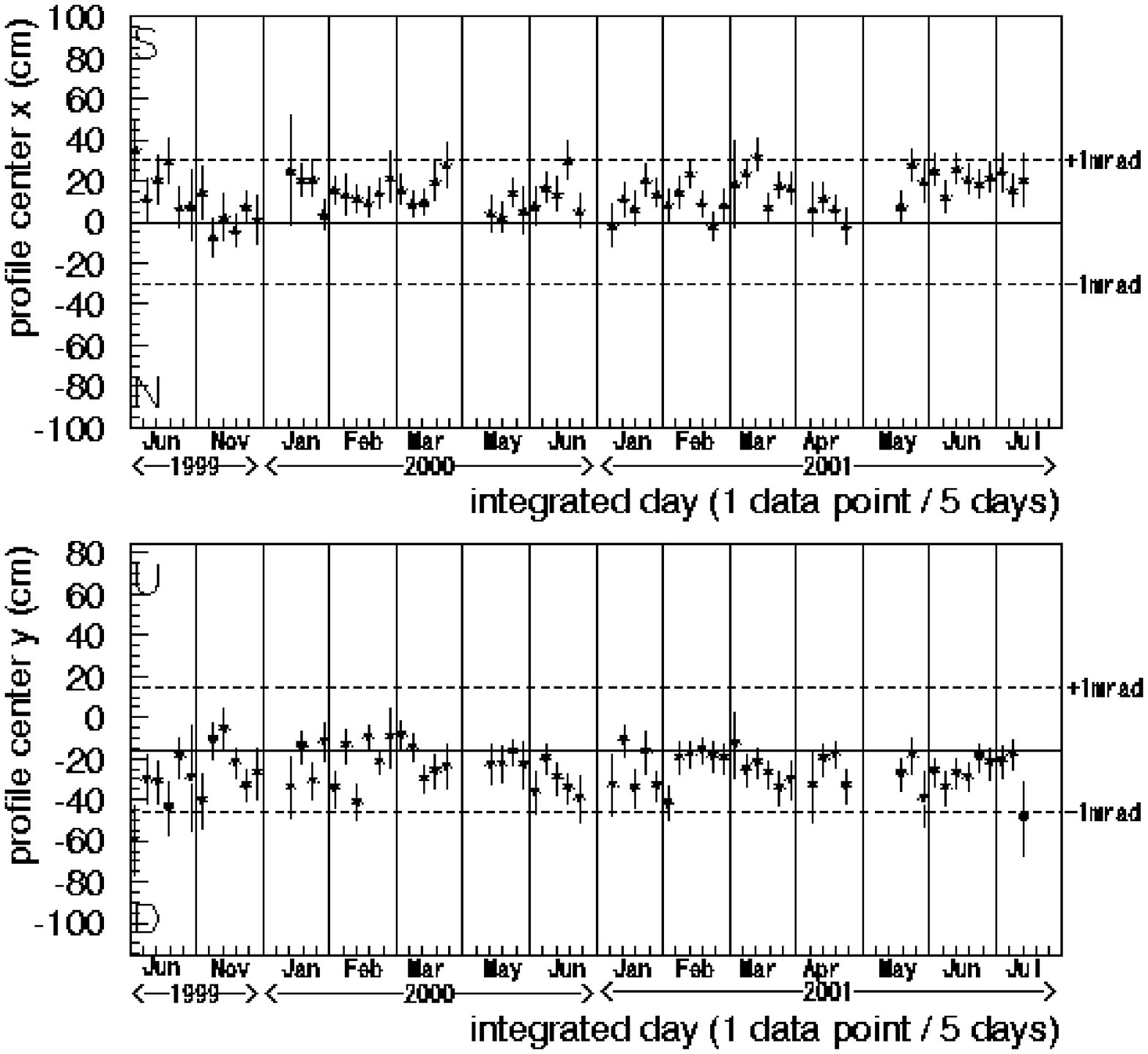}}
}
\center{
\caption{
Time variations of the beam directions measured by MUMON (left)
and the time variations of the profile center measured by the vertex
distributions in MRD (right). 
The top figures and the bottom figures are the horizontal distributions and
the vertical distributions, respectively.
The SK direction and 1~mrad off axis are shown by the solid and dashed lines,
respectively.
}}
\label{direction}
\end{figure}

The neutrino beam direction is also measured using neutrino interactions
in MRD\cite{mrd}. From spatial constructions of the vertex positions, the center
of the vertex positions is found to agree with the SK direction within 1~mrad.
The time variation of the beam center, also plotted in Fig.~2,
shows that the steering of the beam direction is stable,
and is consistent with the results from the muon monitors.

Since the angular acceptance of the SK detector
from the KEK site is about 0.2~mrad, and
the energy spectrum of the neutrino beam is expected to be
uniform within 3~mrad from the center of
the beam axis, the adjustment of the neutrino beam
direction ($<$ 1~mrad) is sufficient.

\begin{figure}[b!]
\center{{\includegraphics[height=7.0cm]{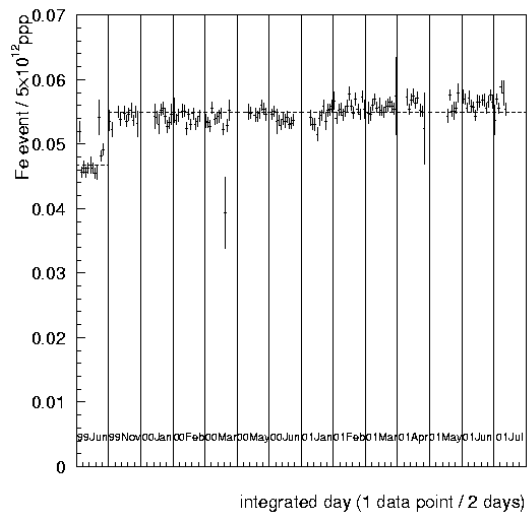}}
\caption{
Time variation of the neutrino event numbers in 1KT.
The June 99 data is smaller than the other periods because
the current of the magnetic horn was different.
The neutrino beam intensity is stable within
the statistical errors.
}}
\label{fig2}
\end{figure}

\subsection{Neutrino beam intensity and its stability}

The absolute neutrino event rates in the FDs are used to calculate
the expected number of events in SK.
The neutrino events in 1KT are reconstructed from the Cherenkov ring pattern
by the same method as that used in the SK atmospheric neutrino analysis.
We selected events with the reconstructed vertex being inside 25.1~tons
of the fiducial volume, which is defined by a 2~m radius, 2~m long cylinder
along the beam axis. Since the average number of neutrino events in the whole
1KT detector is about 0.2~events/spill, multiple neutrino events in one spill
must be taken into consideration. The total sum of PMT signals, which are recorded by
FADC, is used for counting the number of neutrino interactions
in a spill. By considering this correction, the number of neutrino
interactions in the fiducial volume was obtained to be $\sim$0.02 events per spill.
This measurement has a 4\% systematic
uncertainty, which is mainly due to vertex resolution of the reconstructions.

Although the absolute event numbers were also studied by SFT and MRD,
the neutrino events in 1KT are
preferable for an absolute event number analysis,
because the possible systematic errors inherent to the
detection method are the same as that of SK, and are canceled in the analysis.
Therefore, details of the SFT and MRD analysis are not
presented in this report.

The stability of the beam intensity is contaneously measured from
the neutrino event rate in MRD because of their large statistics.
A detailed description of the event reconstruction can be found in \cite{mrd}. 
The time variation of the neutrino interactions in MRD
is shown in Fig.~3.
The neutrino beam intensity is stable within a few \%.

\subsection{$\nue/\num$ ratio}
The $\nue/\num$ ratio of the neutrino beam at the KEK site
was measured by 1KT and FGD.
The idea of the measurements is given in \cite{moriond}.
An analysis with 1KT is still under way, and no numerical result has been
obtained yet.
On the other hand, a very preliminary result with 
FGD is reported to be
(1.8$\pm$0.6\hbox{${{+0.8}\atop{-1.0}}$})\%\cite{Yoshida},
where the expectation based on a Monte-Carlo simulation is 1.3\%.
The original neutrino beam has been proved to be almost
a pure $\num$ beam.

\subsection{Neutrino energy spectrum}

The neutrino energy spectrum was examined with 1KT and FGD.
The 1KT has a high efficiency for low-energy neutrinos below 1~GeV
and a full 4$\pi$ coverage in the solid angle. However, the efficiency for
muons above $\sim$1~GeV is poor because most of the muons escape from the detector.
On the contrary, although FGD has a high efficiency for higher momentum muons,
the efficiency for low-momentum muons is relatively poor, and
the angular acceptance is limited to be within 50 degrees from the beam axis.
By combining the 1KT and FGD results, the entire
range of the neutrino energy spectrum can be covered.

In the 1KT detector, events with only one $\mu$-like
Cherenkov ring are selected.
It was also required that the vertex position is in the 25.1~tons of
the fiducial volume described in the previous section,
and the particles are fully-contained in the inner detector.
To determine the neutrino energy, single $\mu$-like
ring events are assumed to be the charged-current quasi-elastic (CCQE)
interaction of muon neutrinos, $\nu_{\mu}N \rightarrow \mu N'$.
Most of the neutrino energy is transfered to the muons in
this interaction mode.
The neutrino energy $(E_{\nu})$ can be directly calculated from the momentum $(p_{\mu})$
and angle $(\theta_{\mu})$ of the outgoing muons as
\begin{equation}
E_{\nu}={{m_{N} E_{\mu}-m^{2}_{\mu}/2}\over{m_{N}-E_{\mu}+p_{\mu} \cos\theta_{\mu}}},
\end{equation}
where $m_{N}$ and $m_{\mu}$ are the masses of the nucleon and the muon, respectively, and
$E_{\mu}=\sqrt{p_{\mu}^{2}+m_{\mu}^{2}}$.
The muon momentum and angle from the beam axis are shown in
Fig.~4 together with the Monte Carlo expectations described below.
 
\begin{figure}[b!]
\center{{\includegraphics[height=7.0cm]{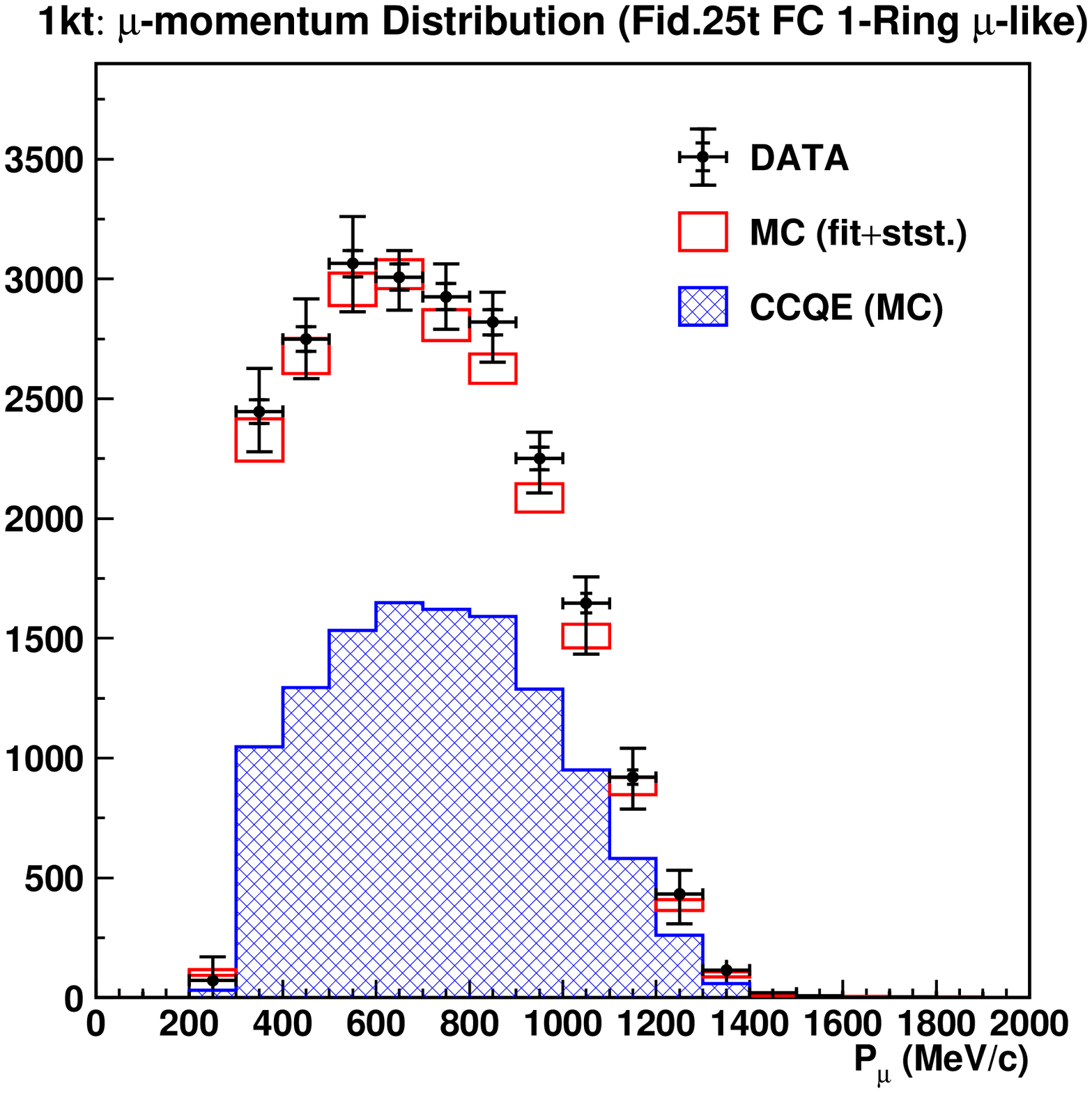}}
\hskip 5mm
        {\includegraphics[height=7.0cm]{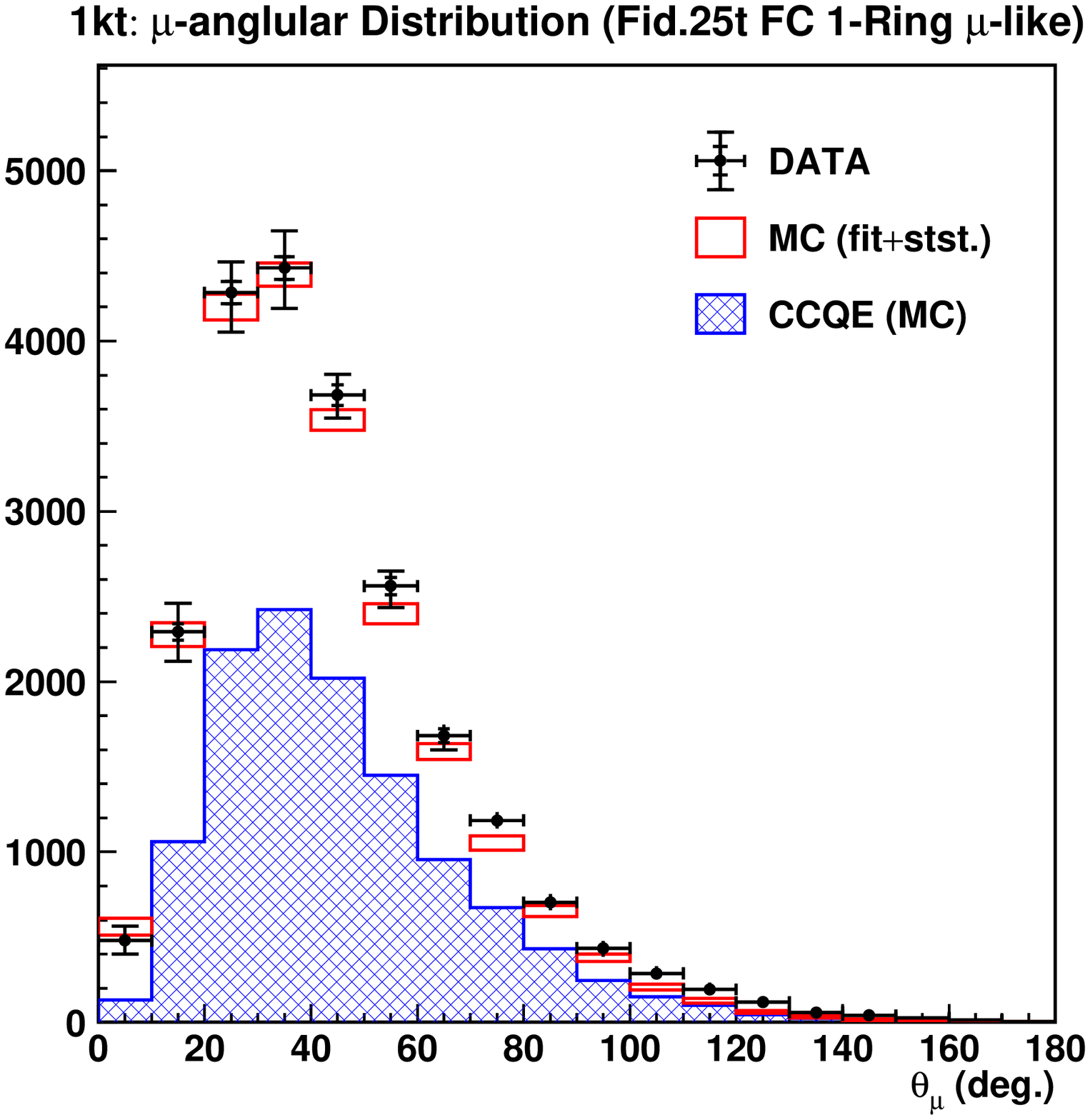}}
}
\center{
\caption{
Muon momentum and angle from the beam axis at the 1KT detector.
The cross is data and the box is a Monte-Carlo simulation with the
best-fit parameters.
}}
\label{ktmomang}
\end{figure}

\medskip

In FGD, neutrino events which satisfy the following criteria are
selected: (1)the vertex position is inside of the 5.94~ton
SFT fiducial volume; (2)one track passes through 3 layers of
scintillating fibers, TRG, LG, and is stopped in MRD.
CCQE events are recognized as 1-track or 2-track events
because the recoiled proton track can be
occasionally identified as another short track.
If the CCQE interaction is assumed, the travel direction of the
recoiled proton can be expected from the direction of the muon track
by using momentum conservation.
The angle between the expected direction
and the real second track $(\Delta\Theta)$ can be used to distinguish CCQE events
from other interactions.
Three categories of events are defined as: (1)1-track events,
(2)2-track events with $\Delta\Theta < 25^{\circ}$, and 
(3)2-track events with $\Delta\Theta > 30^{\circ}$.
The second and the third samples are CCQE and non-CCQE enriched event
samples, respectively.
The momentum $(p_{\mu})$ and angle $(\theta_{\mu})$ distributions
for three events samples are shown in Fig.~5.

\begin{figure}[b!]
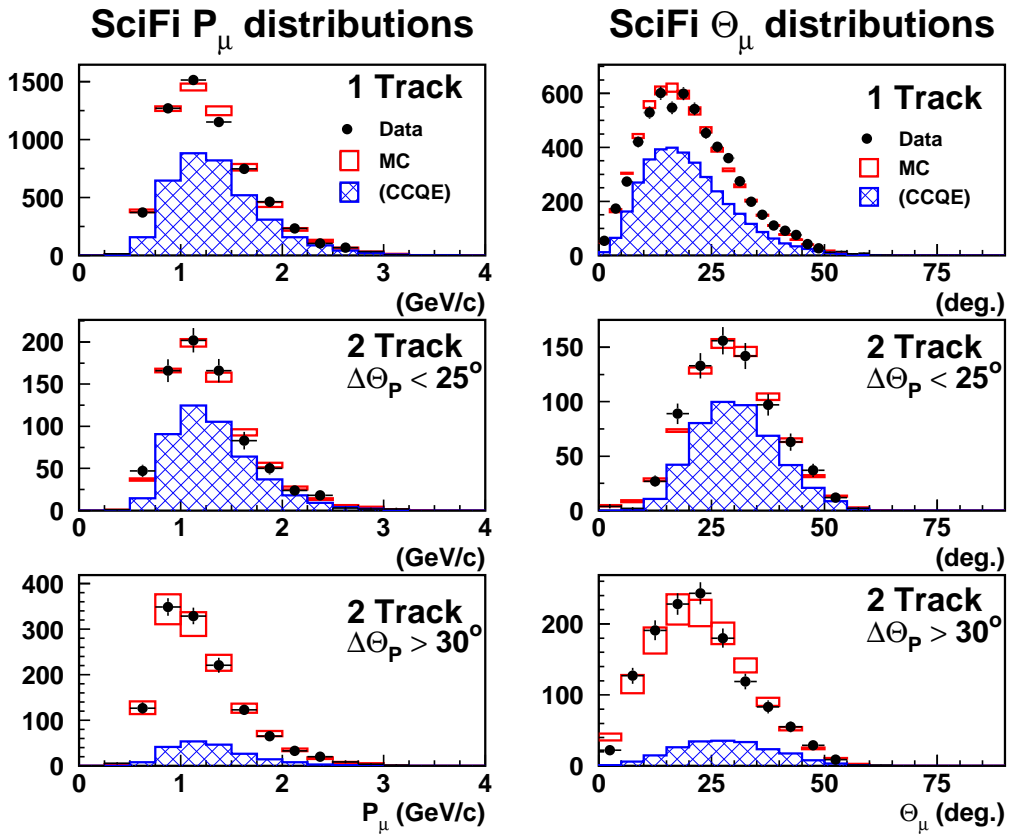

\center{{\includegraphics[height=11.0cm]{sfmom.epsi}}
\hskip 5mm
        {\includegraphics[height=11.0cm]{sfang.epsi}}
}
\caption{
Muon momentum and angle from the beam axis at the FGD detector.
The cross is data and the box is a Monte-Carlo simulation with the
best-fit parameters.
}
\label{sft}
\end{figure}

The expectations are based on GEANT\cite{GEANT} with a detailed description of 
the materials and magnetic fields in the target region and the decay volume.
It uses measurements of the primary-beam intensity
and profile at the target as input parameters.
The primary proton interactions on aluminum are modeled with
a parameterization of hadron production
data\cite{hadr}. Other hadronic interactions are treated by 
GEANT-CALOR\cite{GCALOR}.

The momentum distribution in 1KT and FGD can be used to determine
the neutrino energy spectrum by the following procedure.
The neutrino energy spectrum is divided into 8 energy regions,
and the corresponding muon momentum distributions are prepared
by a Monte-Carlo simulation for each detector. 
The weighting factor of each region as well as other systematic errors are assumed
to be fitting parameters. By the comparing with the real data in 1KT and FGD,
the neutrino energy distribution is determined.
The most probable values and their possible ranges of the systematic errors
are also obtained from the fitting, and are applied to the following SK analysis.

The best-fit results are also shown in Fig.~5 and Fig.~6.
The agreement with the data is excellent. 

\subsection{Extrapolation of the neutrino flux}
 
\begin{figure}[b!]
\center{\includegraphics[height=9.0cm]{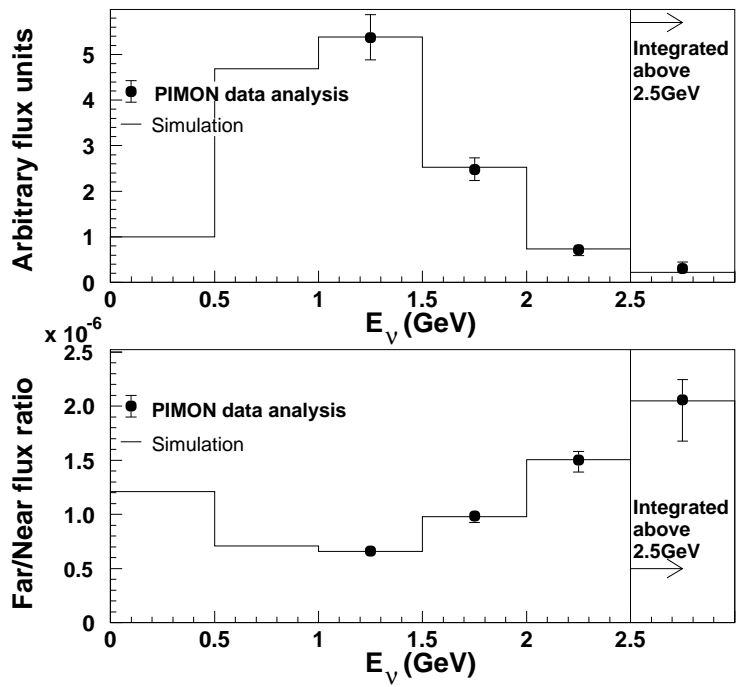}
\caption{
The top figure shows the spectrum shape of the muon energy at FDs.
The bottom figure shows the far to near $\num$ flux ratio.
The histograms are from the beam simulations.
The data points are derived from the PIMON measurements.
}}
\label{pimon}
\end{figure}

The production of neutrinos is calculated from the simple two-body decay
of the charged pions. Therefore, 
once the kinematic distribution of charged pions after the second
magnetic horn is known, it is possible to predict the muon neutrino
spectrum at any distance.

The pion momentum and angular distribution are measured by PIMON,
whose characteristics are given in Table~2.
In PIMON, intensity and shape of the Cherenkov image in the focal plane
are measured by the PMT array. The measurement is a superposition
of slices of the Cherenkov ring from charged pions of various momentum
and angles.  
The refraction indices are controlled by the gas pressure,
and seven independent measurements at different Cherenkov thresholds
are done. From these measurements, the relative pion beam intensities
are obtained as a function of the momentum and angle.
To avoid contamination of the Cherenkov light from the surviving 12-GeV
primary protons, the refraction indices are adjusted to be below the
Cherenkov threshold of the 12-GeV proton. Accordingly, an analysis for the
pion momentum less than 2~GeV is not possible, and the measurable neutrino
energy range is limited to be larger than 1~GeV.
A more detailed description of the PIMON and the analysis method
can be found in \cite{maruyama}.

Fig.~6 shows the $\num$ energy spectrum shape and the flux ratio in SK
and FDs (so-called Far/Near flux ratio)
obtained by the above-mentioned analysis together with the results
from a Monte-Carlo simulation.
The agreements between this analysis and the simulation are excellent.
Fig.~6(bottom) can be used to calculate the expected event numbers in SK
from the observations in the FDs.

\section{Observation in Super-Kamiokande}

To obtain beam-correlated fully contained neutrino interactions,
an event selection similar to an atmospheric neutrino
analysis\cite{SuperK} was applied.
For the selected events,
the time correlation with the neutrino beam was then examined.
Fig.~7 shows the time difference between the neutrino beam
and the events obtained from atmospheric neutrino selection.

Considering the neutrino beam duration (1.1$\mu$sec)
and accuracy\cite{UTC} of
the absolute time determination ($<0.2\mu$sec), events within a 1.5$\mu$sec
time window covering the neutrino beam period were selected.
A total of 56 fully-contained events were found
in 22.5kt of the fiducial volume.
The number of the neutrino events and the results from the particle
identification\cite{Kasuga} are summarized in Table~4.
Because the expected atmospheric neutrino background in the fiducial
volume within the neutrino beam period was calculated to
be $1.3\times 10^{-3}$ events,
the 56 events in the fiducial volume are a clear signal of neutrinos
from KEK.

\begin{figure}[h!]
\center{\includegraphics[height=4.0cm]{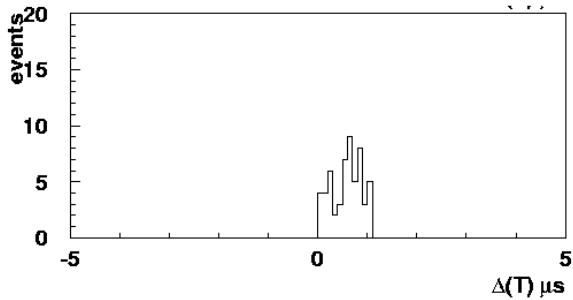}
\caption{
Time correlation between the neutrino beam period and SK events
which are selected by the standard atmospheric neutrino analysis.
Events in the 1.5~$\mu$sec gate ($-0.3~\mu$sec $\sim$ 1.2~$\mu$sec)
are finally selected.
}}
\label{fig4}
\end{figure}

\section{Oscillation analysis}

Strategies of the \nmnt oscillation study at K2K are summarized as follows.
If $\num$ oscillate to $\nut$, the absolute event number of neutrino interactions in
SK is reduced because $\nut$ interact only through neutral current interactions
in the K2K neutrino energy range.
In addition, the neutrino energy spectrum in SK should be
distorted, because the oscillation probability
is a function of the neutrino energy.

In the following two subsections, the status of the data
analyses on the absolute event numbers
and on the distortion of the neutrino energy spectrum are presented.
In the final subsection, an analysis of the constraints on the neutrino oscillation
parameters is discussed.

\subsection{Absolute event numbers}

The expected event numbers in SK are calculated
from the neutrino event rate in FDs, and an extrapolation from the FDs to SK.
The numbers obtained in 1KT, SFT, and MRD are used
for the event rate in the FDs, as reported in ${\it 2.2}$. 
The Far/Near flux ratio 
has already been obtained from PIMON, as presented in ${\it 2.5}$.
The expectations based on the data from the FDs
are 80.1\hbox{${{+6.2}\atop{-5.4}}$}(1KT),
87.5\hbox{${{+10.6}\atop{-11.9}}$}(SFT), and
87.4\hbox{${{+12.7}\atop{-13.9}}$}(MRD).
These results are consistent with each other.
As mentioned in ${\it 2.2}$,
we used the numbers from 1KT as official numbers
because of its small systematic errors.  

\begin{table}[t!]
\caption{
Number of neutrino events in SK. Expectations based on the
event rate at 1KT, SFT, and MRD are also shown.
}
\begin{center}
\begin{tabular}{lrrrr}
\hline
\hline
Event Category~~~~~~~~~~~~& SK data~~~~~~ && Expected & \\
                          &         &(1KT)~~~~&(SFT)~~~~&(MRD)~~~~\\
\hline
Single ring events        & 32~~~~~~~~~&&& \\
~~($\mu$-like)            & 30(29)~~   &&& \\
~~(e-like)                &  2~~~~~~~~ &&& \\
Multi ring events         & 24~~~~~~~~~&&& \\
\hline
Total                        & 56~~~~~~~~~& 80.1\hbox{${{+6.2}\atop{-5.4}}$}~~~
 & 87.5\hbox{${{+10.6}\atop{-11.9}}$} & 87.4\hbox{${{+12.7}\atop{-13.9}}$} \\
\hline
\hline
\end{tabular}
\end{center}
\end{table}

The statistical probability that the observation is equal
to or smaller than 56, where the expectation is
80.1\hbox{${{+6.2}\atop{-5.4}}$}, is about 1\%.
The observation is significantly smaller than the expectation.

\begin{figure}[b!]
\center{\includegraphics[height=8.0cm]{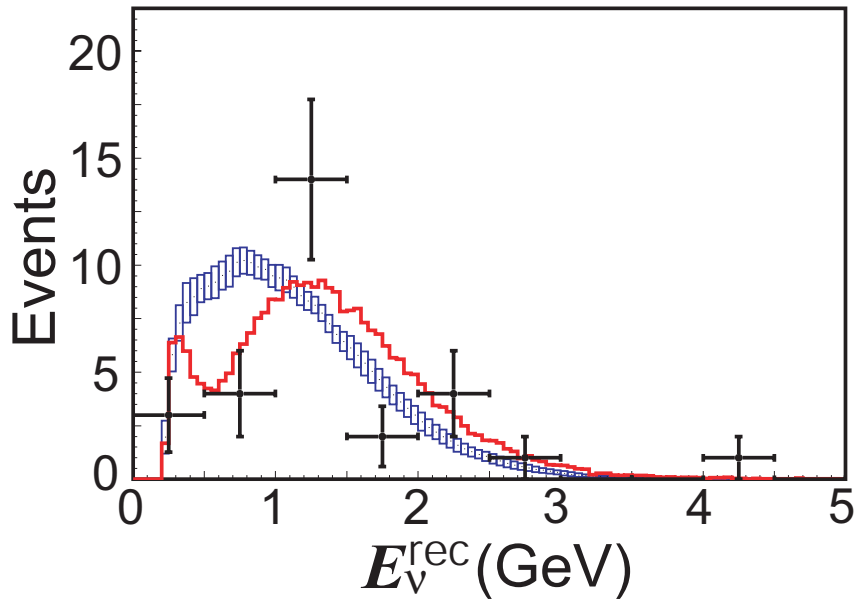}
\caption{
Neutrino energy spectrum reconstructed from 29 single $\mu$-like
events in SK shown by points with error bars.
The blue boxes are the expectations in the case of null oscillation.
The sizes of the boxes are the systematic errors of the expectations.
The red histogram is the expectation for the best-fit oscillation parameters;
$(\Delta m^{2}, \sin^{2}2\theta) = (2.8\times 10^{-3}{\rm eV^{2}},1).$ 
}}
\label{enusk}
\end{figure}

\subsection{Shape of the neutrino energy spectrum}

The neutrino energy spectrum is calculated from 29 single-ring $\mu$-like events.
(Among 30 single-ring $\mu$-like events listed in Table~4, one event recorded in the early
stage of the experiment is not used in the spectrum analysis because
the condition of the neutrino beam line is different from that of
the other data-taking period. See also Fig.~3.)

Fig.~8 shows the energy spectrum of the 29 events with expectations.
These expectations were obtained from the neutrino energy spectrum discussed
in ${\it 2.4}$, and the Far/Near flux ratio presented in ${\it 2.5}$.
A reduction of the neutrino event in the 0.5-1.0~GeV bin is obvious.
If the absolute normalization of the flux is assumed to be
a free parameter without any constraints, the probability that the
data is explained by statistical fluctuation of the null oscillation
is calculated to be about 15\%.
A more quantitative analysis is done by combining with the absolute
event numbers.

\subsection{Constraints on the oscillation parameters}

To evaluate the absolute number of events and the spectrum shape simultaneously,
an un-binned maximum likelihood analysis is employed.
The likelihood is defined as the product of the normalization term and
the shape term:
\begin{equation}
{\cal L}(\Delta m^{2},\sin^{2}2\theta)={\cal L}_{norm}(\Delta m^{2},\sin^{2}2\theta)
\times{\cal L}_{shape}(\Delta m^{2},\sin^{2}2\theta).
\end{equation}
The normalization term, ${\cal L}_{norm}$, is a simple  poisson probability of observing
$N_{obs}$(=56) for the prediction $N_{exp}(\Delta m^{2}, \sin^{2}2\theta)$.
The shape term, ${\cal L}_{shape}$, is the product of the probabilities for each
single $\mu$-like events:
\begin{equation}
{\cal L}_{shape}(\Delta m^{2},\sin^{2}2\theta) =
\prod_{i=1}^{N_{1\mu}} p_{i}(E_{i};\Delta m^{2},\sin^{2}2\theta); 
\end{equation}
where $p_{i}$ is the probability of $i$-th event having
energy $E_{i}$ at a given set of oscillation parameters.
$N_{1\mu}$ is the number of single $\mu$-like events used in the analysis,
and is 29.

When the likelihood, ${\cal L}$, is maximized, various systematic errors
are taken into consideration. The source of the systematic errors include
the spectrum shape, the Far/Near flux ratio, non-CCQE/CCQE ratio,
the efficiency in SK, the absolute energy scale in SK and so on.
Not only the individual systematic errors, but also
the correlation between them are carefully taken into account.
For the systematic error parameters which are common with FD,
their probable range from the FD analysis are considered.
More details about the systematic errors and their threatments are not discussed in
this report, but is reported are \cite{k2kosc}. 

\begin{figure}[b!]
\center{\includegraphics[height=10.0cm]{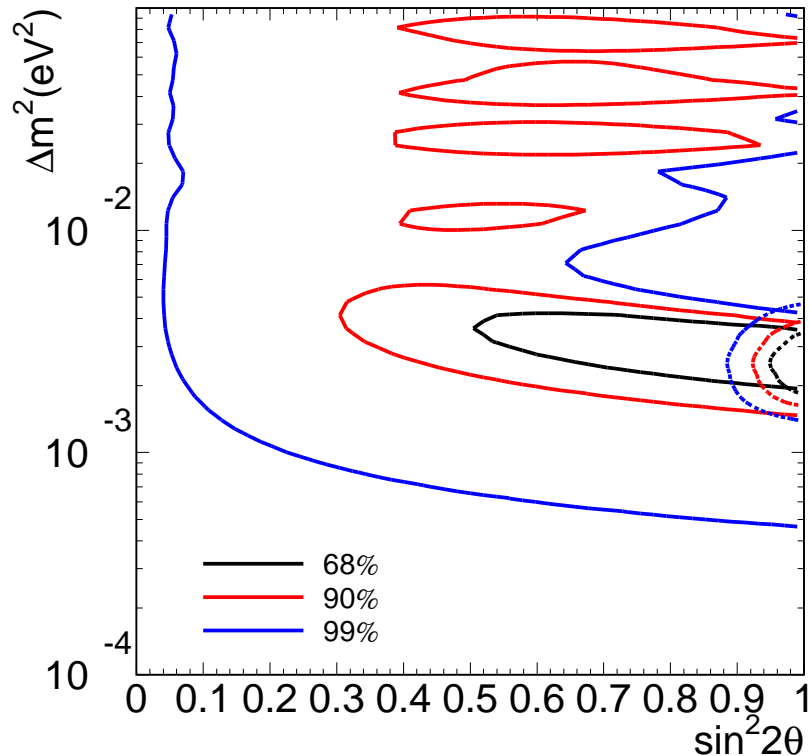}
\caption{
The 68\%, 90\% and 99\% C.L. allowed region on the neutrino oscillation parameters
for \nmnt oscillation.
The K2K results (solid lines) are overlaid with the SK atmospheric neutrino
results (dashed lines). 
}}
\label{contour}
\end{figure}

The oscillation parameter set which maximize
the likelihood, ${\cal L}$, was calculated to be
\begin{equation}
(\Delta m^{2}, \sin^{2}2\theta) = (2.8\times 10^{-3}{\rm eV^{2}}, 1.0).
\end{equation}
In this case, the expected total event number is 54.2.
The expected energy spectrum is shown by the red line in Fig.~8.
The agreements with the experimental results are excellent.

The probability that the K2K results can be explained by a statistical
fluctuation of null oscillation is calculated from the likelihood ratio
of the null oscillation to the best-fit point.
The results were obtained to be 0.4\% or 0.7\%, depending on
the two independent treatments of the systematic errors in the likelihood\cite{k2kosc}.
We conclude that the null oscillation probability is less than 1\%.

Finally, constraints on the oscillation parameters are studied from the
likelihood ratio to the best fit. The allowed region in the oscillation
parameter plane is shown in Fig.~9. The 90\% C.L. allowd region
on $\Delta m^{2}$ for the full mixing, $\sin^{2}2\theta=1$, is
\begin{equation}
\Delta m^{2} = (1.5 \sim 3.9)\times 10^{-3}{\rm eV^{2}}.
\end{equation}
These results agree well with the most recent SK atmospheric neutrino
analysis\cite{kajita}, as shown in Fig.~9.

\section{Summary}

The K2K long-baseline neutrino-oscillation experiment
has been successfully operated since 1999.
By the end of 2001, a total intensity of 47.9$\times 10^{18}$
protons on target was accumulated.
A total of 56 fully-contained neutrino
interactions in the 22.5kt of the fiducial volume of the
SK detector were observed,
where the expectation based on the data from the Front Detectors
is 80.1\hbox{${{+6.2}\atop{-5.4}}$}.
The neutrino energy spectrum obtaind from 29 single $\mu$-like ring events
shows a clear deficit in neutrino energy between 0.5~GeV and 1.0~GeV.
The probability that the measurements at SK can be explained by a statistical
fluctuation of the null oscillation is less than 1\%.
The 90\% C.L. allowed region on the oscillation parameters is 
$\Delta m^{2}=(1.5\sim 3.9)\times 10^{-3}$eV$^{2}$
for $\sin^{2}2\theta=1$. This result is consistent with the results from the SK
atmospheric neutrino analysis.

\end{document}